\documentclass[submission,copyright,creativecommons]{eptcs}
\providecommand{\event}{ML 2017} 
\usepackage{underscore}          
\usepackage{amsmath,xcolor,listings,url,quoting,cprotect}
\hypersetup{
    colorlinks=false,
    pdfborder={0 0 0},
}
\definecolor{darkgreen}{rgb}{0,0.2,0}
\definecolor{darkpurple}{rgb}{0.3,0.1,0.3}

\lstdefinelanguage{OCaml}{
  keywords={
    and,as,assert,asr,begin,class,constraint,do,done,downto,effect,else,end,exception,
    external,false,for,fun,function,functor,if,implicit,in,include,inherit,initializer,
    land,lazy,let,lor,lsl,lsr,lxor,macro,match,method,mod,module,mutable,new,object,
    of,open,or,private,rec,sig,struct,then,to,true,try,type,val,virtual,when,
    with,while},
  comment=[s]{(*}{*)},
}

\lstdefinestyle{text}{
  basicstyle=\ttfamily\small,
  basewidth=0.5em,
}

\lstdefinestyle{OCaml}{
  basicstyle=\ttfamily\small,
  basewidth=0.5em,
  commentstyle=\color{darkgreen},
  escapeinside={(**}{)},
  keywordstyle=\color{darkpurple}\bfseries,
  language=OCaml,
  stringstyle=\color{blue},
  showstringspaces=false,
  mathescape=true,
  moredelim=**[is][]{?}{?},
  moredelim=**[is][]{&}{&},
}
\title{First-Class Subtypes}
\author{
Jeremy Yallop
\institute{University of Cambridge}
\email{jeremy.yallop@cl.cam.ac.uk}
\and
Stephen Dolan
\institute{University of Cambridge}
\email{stephen.dolan@cl.cam.ac.uk}}

\def\titlerunning{First-Class Subtypes}
\def\authorrunning{J. Yallop \& S. Dolan}
\begin{document}
\maketitle

\begin{abstract}

First class type equalities, in the form of generalized algebraic data
types (GADTs), are commonly found in functional programs.
However, first-class representations of other relations between types,
such as subtyping, are not yet directly supported in most functional
programming languages.

We present several encodings of first-class subtypes using existing
features of the OCaml language (made more convenient by the proposed
modular implicits extension), show that any such encodings are
interconvertible, and illustrate the utility of the encodings with 
several examples.
\end{abstract}

\section{Introduction}

One appealing feature of ML-family languages is the ability to define
fundamental data structures --- pairs, lists, streams, and so on
--- in user code.
For example, although \textit{lazy} computations are supported as a
built-in construct in OCaml, it is also possible to implement them as
a library.

\paragraph{Laziness in variants}

The following data type can serve as a basis for lazy computations\footnote{
The code in this paper uses the proposed \textit{modular implicits} extension to OCaml~\cite{modular-implicits}.
The modular implicits compiler
can be installed as the \textsc{opam} switch \texttt{4.02.0+modular-implicits}.
}:

\begin{center}\begin{tabular}{c}\begin{lstlisting}
type 'a lazy_cell = | Thunk of (unit -> 'a)
                    | Value of 'a
                    | Exn of exn
\end{lstlisting}\end{tabular}\end{center}

\noindent
The constructors of a \lstinline{lazy_cell} value represent
  the three possible states of a lazy computation:
   it may be
     an unevaluated thunk,
     a fully-evaluated value, or
     a computation whose evaluation terminated with an exception.
Since the state of a lazy computation may change over time,
  lazy values are represented
    as mutable references
      that hold \lstinline{lazy_cell} values:

\begin{center}\begin{tabular}{c}\begin{lstlisting}
type 'a lzy = 'a lazy_cell ref
\end{lstlisting}\end{tabular}\end{center}

\noindent
Finally, there are two operations:
  \lstinline{delay} creates a thunk from a function, while
  \lstinline{force} either
    forces a thunk and caches the result,
    or returns the value or exception cached by a previous call.

\begin{center}\begin{tabular}{cp{0.4in}c}\begin{lstlisting}
let delay f = ref (Thunk f)





$~$
\end{lstlisting}&&
\begin{lstlisting}
let force r = match !r with
  | Thunk f ->
     (match f () with
      | v -> r := Value v; v
      | exception e -> r := Exn e; raise e)
  | Value v -> v
  | Exn e -> raise e
\end{lstlisting}\end{tabular}\end{center}

\paragraph{Laziness invariants}

The characterising feature of lazy computations is that each
computation is run only once, although the result may be read many
times.
The \lstinline{delay} and \lstinline{force} functions enforce this
property. For simplicity, we assume that lazy computations do not
invoke themselves recursively. (More sophisticated implementations of
laziness generally include a fourth state \lstinline{In_progress} to
guard against this occurrence).

Concealing the representation type of \lstinline{lzy} behind a module
interface ensures that other parts of the program cannot violate this
invariant:

\begin{center}\begin{tabular}{cp{0.4in}c}\begin{lstlisting}
module Lzy :
sig
  type 'a t
  val delay : (unit -> 'a) -> 'a t
  val force : 'a t -> 'a
end
\end{lstlisting}&&
\begin{lstlisting}
module Lzy =
struct
  type 'a t = 'a lzy
  let delay = $\ldots$ (* as above *)
  let force = $\ldots$ (* as above *)
end
\end{lstlisting}\end{tabular}\end{center}

\paragraph{Laziness invariance}

This simple implementation has the same behaviour as the built-in
\lstinline{lazy} type.
However, there is one notable difference: unlike the built-in type,
our \lstinline{Lzy.t} is not \textit{covariant}.

Covariant types are parameterised types that preserve the subtyping
relationships between parameters.  For example, if \lstinline{u} is a
subtype of \lstinline{v} then, because \lstinline{option} is covariant,
\lstinline{u option} is a subtype of \lstinline{v option}.
In OCaml, types may be marked as covariant by adding a \lstinline{+}
before a type parameter.
For example, here is a definition of the covariant type \lstinline{option}:
\begin{center}\begin{tabular}{c}\begin{lstlisting}
type +'a option = None | Some of 'a
\end{lstlisting}\end{tabular}\end{center}

Not every type can be marked as covariant in this way.  For example,
it would not be safe to allow \lstinline{ref}, the type of mutable
cells, to behave covariantly.
If \lstinline{ref} were covariant then a program could coerce a value
of type \lstinline{u ref} to \lstinline{v ref}, and then store a value
of another subtype of \lstinline{v}, unrelated to \lstinline{u}, in
the cell.
OCaml therefore prohibits covariance for type parameters that appear
under \lstinline{ref}, as in the definition of \lstinline{lzy}.

The lack of covariance in our \lstinline{Lzy.t} has two significant
consequences for programmers.
First, computations constructed using the built-in type can be
coerced, while computations constructed using our \lstinline{Lzy.t}
cannot.
Here is a coercion that eliminates a method \lstinline{m} from the
type of a built-in lazy computation:

\begin{center}\begin{tabular}{c}\begin{lstlisting}
# let o = object method m = () end;;
val o : < m : unit > = <obj>
# (lazy o :> < > Lazy.t);;
- : <  > Lazy.t = <lazy>
\end{lstlisting}\end{tabular}\end{center}

\noindent
An attempt to similarly coerce our \lstinline{Lzy.t} fails:

\begin{center}\begin{tabular}{c}\begin{lstlisting}
# (Lzy.delay (fun () -> o) :> < > Lzy.t);;
Characters 0-40:
  (Lzy.delay (fun () -> o) :> < > Lzy.t);;
  ^^^^^^^^^^^^^^^^^^^^^^^^^^^^^^^^^^^^^^^^
Error: Type < m : unit > Lzy.t is not a subtype of <  > Lzy.t 
       The second object type has no method m
\end{lstlisting}\end{tabular}\end{center}

Second, \lstinline{let}-bound computations constructed using the
built-in \lstinline{lazy} receive polymorphic types, following the
\textit{relaxed value restriction}, which generalizes type variables
that appear only in covariant
positions~\cite{relaxing-the-value-restriction}:

\begin{center}\begin{tabular}{c}\begin{lstlisting}
# let f = Lazy.from_fun (fun () -> []);;
val f : 'a list Lazy.t = <lazy>
\end{lstlisting}\end{tabular}\end{center}

\noindent
In contrast, types built using our \lstinline{Lzy.t} are
ungeneralized, as the leading underscore on the type variable \lstinline{'_a} indicates:\footnote{See \url{https://caml.inria.fr/pub/docs/manual-ocaml/polymorphism.html}}

\begin{center}\begin{tabular}{c}\begin{lstlisting}
# Lzy.delay (fun () -> []);;
- : '_a list Lzy.t = <abstr>
\end{lstlisting}\end{tabular}\end{center}

The interface to \lstinline{Lzy.t} only exposes read operations, and
so it would be safe for the type to be treated as covariant in its
parameter.
However, the assignment of variance considers only the use of the
parameter in the definition of \lstinline{Lzy.t}, not the broader
module interface.
Since the type parameter is passed to the the invariant type
\lstinline{ref} of mutable cells, the type \lstinline{Lzy.t} is also
considered invariant.


These shortcomings in the \lstinline{Lzy} interface can be overcome
with more flexible treatment of subtyping and variance\cprotect\footnote{%
An alternative solution is to switch to a higher-order representation
of lazy computations:
\begin{lstlisting}[basicstyle=\ttfamily\footnotesize]
type 'a t = unit -> 'a
let delay f =
  let r = ref (Thunk f) in
  fun () -> match !r with
    | Thunk f ->
       (match f () with
        | v -> r := Value v; v
        | exception e -> r := Exn e; raise e)
    | Value v -> v
    | Exn e -> raise e
let force f = f ()
\end{lstlisting}  
With this representation the type \lstinline{t} is covariant in its
parameter, which occurs only in a positive position.  However, the
higher-order representation makes lazy values more difficult to inspect.}.
In particular, making subtypes \textit{first-class} makes it possible
to tie together the type definition and the functions exposed in the
interface in the consideration of variance assignment, and so make
\lstinline{Lzy.t} covariant.

Variants of first-class subtypes may be found in advanced type systems
in the research literature, such as
Cretin and R{\'e}my's F$_\iota$~\cite{cretin:hal-00650910}.
Our contribution here is to show that first-class subtypes can be
\textit{encoded} using the features of an existing real-world
functional programming language.

First-class subtypes can be defined using a binary type constructor:

\begin{center}\begin{tabular}{c}\begin{lstlisting}
type (-'a, +'b) sub
\end{lstlisting}\end{tabular}\end{center}

\noindent
that has a single constructor:

\begin{center}\begin{tabular}{c}\begin{lstlisting}
val refl : ('a, 'a) sub
\end{lstlisting}\end{tabular}\end{center}

\noindent
and an operation that turns a \lstinline{sub} value into a function:

\begin{center}\begin{tabular}{c}\begin{lstlisting}
val (>:) : 'a -> ('a, 'b) sub -> 'b
\end{lstlisting}\end{tabular}\end{center}

These three elements, considered in more detail in
Section~\ref{section:definition} are sufficient to define a covariant
variant of \lstinline{Lzy}.
Section~\ref{section:definition} adds an additional constuctor,
\lstinline{lift} that, together with the three elements above,
suffices as a basis to define a range of useful subtyping operations.


Here is a second, covariant interface to lazy computations:

\begin{lstlisting}
module CovLzy :
sig
  type +'a t
  val delay : (unit -> 'a) -> 'a t
  val force : 'a t -> 'a
end
\end{lstlisting}

The implementation of \lstinline{CovLzy} can be constructed from
the combination of \lstinline{Lzy} and first-class subtypes.
Here is the definition of \lstinline{CovLzy.t}

\begin{center}\begin{tabular}{c}\begin{lstlisting}
type +'b t = L : 'a Lzy.t * ('a, 'b) sub -> 'b t
\end{lstlisting}\end{tabular}\end{center}

\noindent
That is, a value of type \lstinline{'a CovLzy.t} is a pair of
a lazy computation of type \lstinline{'a Lzy.t}
and a value of type \lstinline{('a, 'b) sub}
that supports coercions from \lstinline{'a} to \lstinline{'b}.

Now \lstinline{CovLzy.delay} builds on \lstinline{Lzy.delay},
pairing a lazy computation with a \lstinline{sub} value\footnote{%
A reviewer observes that the permission-oriented language
Mezzo~\cite{mezzo} uses a related approach. In Mezzo, a witness that
\lstinline{s} is a subtype of \lstinline{t} can be encoded as a
permission of the form \lstinline{id @ a -> b}, which can be read as
``the identity function has type \lstinline{a -> b}'', and these
witnesses are also used in Mezzo to make the \lstinline{lazy} type
covariant:
\url{http://protz.github.io/mezzo/code_samples/lazy.mz.html}.}:

\begin{center}\begin{tabular}{c}\begin{lstlisting}
let delay f = L (Lzy.delay f, refl)
\end{lstlisting}\end{tabular}\end{center}

Finally, the definition of \lstinline{CovLzy.force} calls
\lstinline{Lzy.force} and applies a coercion to the value returned:

\begin{center}\begin{tabular}{c}\begin{lstlisting}
  let force (L (sub, l)) =
    match Lzy.force l with
    | v -> (v >: sub)
    | exception e -> raise e
\end{lstlisting}\end{tabular}\end{center}

These additions to make \lstinline{CovLzy} covariant bring its
behaviour closer to the behaviour of the built-in \lstinline{lazy}.
For example, \lstinline{let}-bound values built by
\lstinline{CovLzy.delay} receive polymorphic types:

\begin{center}\begin{tabular}{c}\begin{lstlisting}
# CovLzy.delay (fun () -> []);;
- : 'a list CovLzy.t = <abstr>
\end{lstlisting}\end{tabular}\end{center}

\section{First-class subtypes defined}
\label{section:definition}

\subsection{Subtypes \textit{\`a la} Liskov \& Wing}

The first ingredient in a representation of subtyping proofs
 is a definition of subtyping.
Here is Liskov and Wing's
characterisation~\cite{a-behavioural-notion-of-subtyping}:
\begin{quotation}
\noindent
Let $\phi(x)$ be a property provable about objects $x$ of type $T$.
Then $\phi(y)$ should be true for objects $y$ of type $S$
  where $S$ is a subtype of $T$.
\end{quotation}

\noindent
For instance,
   properties of a record type $r$
   should also hold for a widening of $r$,
     since the extra fields can be ignored.
And dually,
   properties of a variant type $v$
   should also hold for a narrowing of $v$:
     any property that holds for all constructors
        also holds for a subset of constructors.

\subsection{Subtypes \textit{\`a la} Curry \& Howard}

The Curry-Howard correspondence
  turns Liskov and Wing's characterisation
    of subtyping
  into an executable program.

With a propositions-as-types perspective~\cite{propositions-as-types},
   a property provable about objects of type $T$
     is represented as a type $\phi(T)$ involving $T$,
   and a proof of that property is a term of that type
\footnote{At least, in total languages. In OCaml the waters are muddied by side-effects and nontermination.}.
Liskov and Wing's proposition that $S$ is a subtype of $T$ corresponds to the following (poly)type:
\[
\forall \phi.\phi(T) \to \phi(S)
\]
\noindent
Two points deserve note.

First, although Liskov and Wing's characterisation is couched in terms
of \textit{objects} $x$ and $y$, it is really about their
\textit{types} $T$ and $S$.
To see this, consider that, in the characterisation, it is sufficient
to know $y$'s type to know that $\Phi(y)$ holds.
Since the characterisation only involves properties about all the
objects of a type, not properties of individual objects, there is no
need for dependent types in the type corresponding to the subtyping
proposition.

Second, a ``property about objects'' is a context that
\textit{consumes} an object.
For example, consider the property ``for every object of type $T$
there is an object of type $R$'',
which can reasonably be described as a property about objects of type $T$,
but not as a property about objects of type $R$.
In a propositions-as-types setting, a proof of this
property is a context that consumes an object of type $T$ and
produces an object of type $R$.
Since the properties of interest are consumers of objects,
$\phi$ ranges over \textit{negative} contexts.

\subsection{Contexts and variance}

\begin{figure}[ht]
\begin{lstlisting}
module type POS = sig type +'a t end
module type NEG = sig type -'a t end
\end{lstlisting}
\begin{lstlisting}
module Id = struct type 'a t = 'a end
\end{lstlisting}
\begin{lstlisting}
module Compose$_{PN}$(F:NEG)(G:POS) = struct type 'a t = 'a F.t G.t end
module Compose$_{PP}$(F:POS)(G:POS) = struct type 'a t = 'a F.t G.t end
\end{lstlisting}%
\caption{\label{figure:negative-contexts}Positive and negative contexts}
\end{figure}

Figure~\ref{figure:negative-contexts} defines
 OCaml signatures, \lstinline{POS} and \lstinline{NEG},
  of positive and negative type contexts.
The \lstinline{-} preceding the type parameter \lstinline{'a}
indicates that {'a} can only appear in negative (contravariant)
positions in instantiations of the signature.
The \lstinline{Id} module
 and \lstinline{Compose} functors 
 represent
   the identity context and
   the composition of two contexts.
Each composition of variance in the argument contexts requires
   a separate \lstinline{Compose}
   (but see \S\ref{section:further-examples} for a generalization).

\subsection{Encoding subtypes}

\begin{figure}[ht]
\begin{lstlisting}
type (-'a, +'b) sub
val refl : ('a, 'a) sub
val lift: {P:POS} -> ('a,'b) sub -> ('a P.t,'b P.t) sub
val (>:) : 'a -> ('a, 'b) sub -> 'b
\end{lstlisting}
\caption{\label{figure:basic-sub-interface}First-class subtypes: minimal interface}
\end{figure}

Figure~\ref{figure:basic-sub-interface}
  defines an interface to subtype witnesses.
A value of type \lstinline{(s, t) sub}
  serves as evidence
   that \lstinline{s} is a subtype of \lstinline{t}.
There are two ways to construct such evidence.
First, \lstinline{refl} represents the fact that
  every type is a subtype of itself.
Second, \lstinline{lift} represents the fact that
  subtyping lifts through covariant contexts,
  which are passed as implicit arguments~\cite{modular-implicits}.
(Lifts through contravariant contexts are defined below in terms
 of the minimal interface.) 
The single destructor, \lstinline{>:},
 which mimics OCaml's built-in coercion operator \lstinline{:>},
 supports converting a value of type \lstinline{s}
 to a supertype \lstinline{t}.

This small interface suffices as a basis for many useful
subtyping-related functions.
For example, the transitivity of subtyping is represented
by a function of the following type:
\begin{center}\begin{tabular}{c}\begin{lstlisting}
  val trans : ('a,'b) sub -> ('b,'c) sub -> ('a,'c) sub
\end{lstlisting}\end{tabular}\end{center}
and may be defined as follows:
\begin{center}\begin{tabular}{c}\begin{lstlisting}
  let trans (type a b c) (x : (a,b) sub) (y : (b,c) sub) =
    let module M = struct type +'d t = (a,'d) sub end
    in x >: lift {M} y
\end{lstlisting}\end{tabular}\end{center}
Here the application \lstinline{lift M y} builds
 a value of type \lstinline{(b M.t, c M.t) sub} ---
 that is to say,
 a value of type \lstinline{((a, b) sub, (a, c) sub) sub} ---
 which is used to coerce \lstinline{x} from type \lstinline{(a, b) sub} to type \lstinline{(a, c) sub}.

Similarly, a function that lifts subtyping witnesses through negative
contexts

\begin{center}\begin{tabular}{c}\begin{lstlisting}
  val lift$_-$ : {N:NEG} -> ('a, 'b) sub -> ('b N.t, 'a N.t) sub
\end{lstlisting}\end{tabular}\end{center}

\noindent
may also be defined by supplying a suitable implementation of
\lstinline{POS} to \lstinline{lift}:

\begin{center}\begin{tabular}{c}\begin{lstlisting}
  let lift$_-$ (type a b) {N:NEG} (x: (a,b) sub) : (b N.t,a N.t) sub =
    let module M = struct type +'b t = ('b N.t, a N.t) sub end in
    refl >: lift {M} x
\end{lstlisting}\end{tabular}\end{center}

Using the variance of \lstinline{sub}, \lstinline{refl} can be used to
define a witness for any subtyping fact that holds in the typing environment.
For example,
in OCaml the object type \lstinline{< m:int >},
with one method \lstinline{m},
is a subtype of the type \lstinline{< >} of objects with no methods.
This fact can be turned into a \lstinline{sub} value by coercing
\lstinline{refl}, either by lowering the contravariant parameter:
\begin{center}\begin{tabular}{c}\begin{lstlisting}
  (refl : (< >, < >) sub :> (<m:int>, < >) sub)
\end{lstlisting}\end{tabular}\end{center}%
or by raising the covariant parameter:
\begin{center}\begin{tabular}{c}\begin{lstlisting}
  (refl : (<m:int>, <m:int>) sub :> (<m:int>, < >) sub)
\end{lstlisting}\end{tabular}\end{center}
The resulting value can be passed freely
 through abstraction boundaries
  that conceal the types involved,
 eventually being used to coerce a value of type \lstinline{<m:int>} to
 its supertype \lstinline{< >}.

The generality of the interface in
Figure~\ref{figure:basic-sub-interface} places constraints on the
implementation.
Most notably,
  since \lstinline{lift} can transport subtyping evidence
   through \textit{any} positive context,
  coercion must pass values through unexamined.
For example,
  \lstinline{lift}
    might be used
     to build a value of type \lstinline{(s list, t list) sub}
     from a value of type \lstinline{(s, t) sub}:
\begin{center}\begin{tabular}{c}\begin{lstlisting}
  let l : (s list, t list) sub = lift {List} s_sub_t
\end{lstlisting}\end{tabular}\end{center}
but applying \lstinline{l}
  cannot
   involve list traversal,
  since
   the subtyping interface
    says nothing about list structure.
A polymorphic interface thus ensures an efficient implementation.


\section{Implementations of subtyping}

\subsection{First-class subtypes via contexts}

\begin{figure}[th]
\begin{lstlisting}
type (-'a, +'b) sub = {N:NEG} -> ('b N.t -> 'a N.t)
let refl {N:NEG} x = x
let lift {P:POS} s {Q:NEG} x = s {Compose$_{PN}$(Q)(P)} x
let (>:) (type b) x f =
  let module M = struct type -'a t = 'a -> b end in
  f {M} id x
\end{lstlisting}
\caption{\label{figure:negative-implementation}First-class subtypes via negative contexts}
\end{figure}

\begin{figure}[th]
\begin{lstlisting}
type (-'a, +'b) sub = {P:POS} -> ('a P.t -> 'b P.t)
let refl {P:POS} x = x
let lift {P:POS} s {Q:POS} x = s {Compose$_{PP}$(P)(Q)} x
let (>:) x f = f {Id} x
\end{lstlisting}
\caption{\label{figure:positive-implementation}First-class subtypes via positive contexts}
\end{figure}

Figure~\ref{figure:negative-implementation} gives an
implementation of Figure~\ref{figure:basic-sub-interface}
based on negative contexts that directly follows Liskov \& Wing's
definition\footnote{
 Edward Kmett has used this approach
  in the \texttt{magpie} library~\cite{kmett-magpie},
 as we discovered after writing this note.
}.
A value of type \lstinline{(s, t) sub} is a proof that \lstinline{t}
can be replaced with \lstinline{s} in any negative context;
operationally it must be the identity, as discussed above,
and so the two constructors \lstinline{lift} and \lstinline{refl} both
correspond to the identity function.
Figure~\ref{figure:positive-implementation} gives a similar but simpler
implementation, based on positive contexts.
Apart from the variance annotations, these definitions mirror the
standard Leibniz encoding of type equality~\cite{first-class-modules-ml-2010}.

\subsection{First-class subtypes as an inductive type}

Consider an ordinary inductive type, say the Peano natural numbers:
\begin{center}\begin{tabular}{c}\begin{lstlisting}
type nat = Zero | Suc nat
\end{lstlisting}\end{tabular}\end{center}
We can write its constructors in the form of a module signature as follows:
\begin{center}\begin{tabular}{c}\begin{lstlisting}
module type NAT = sig
  type t
  val zero : t
  val suc : t -> t
end
\end{lstlisting}\end{tabular}\end{center}
What it means for the type \lstinline{nat} to be inductive is that it
is an \emph{initial algebra} for this signature: first, it implements
the signature by providing \lstinline{Zero} and \lstinline{Suc}, and
secondly, for any other implementation \lstinline{M}, we have a
function mapping \lstinline{nat} to \lstinline{M.t} that maps
\lstinline{Zero} to \lstinline{M.zero} and \lstinline{Suc} to
\lstinline{M.suc}:
\begin{center}\begin{tabular}{c}\begin{lstlisting}
let rec primrec = function
| Zero -> M.zero
| Suc n -> M.suc (primrec n)
\end{lstlisting}\end{tabular}\end{center}

In defining the type \lstinline{nat}, we made use of OCaml's built-in
support for inductive types. Lacking this, we could have used the
initial algebra definition directly, and defined the type
\lstinline{nat} as follows:
\begin{center}\begin{tabular}{c}\begin{lstlisting}
type nat = {M : NAT} -> M.t
\end{lstlisting}\end{tabular}\end{center}

This corresponds to the Church encoding of the natural
numbers~\cite{church-types},
in which a natural number is anything that can produce a \lstinline{M.t} from
\lstinline{M.zero : M.t} and \lstinline {M.suc : M.t -> M.t}.
Here and elsewhere we're using the modular implicits extension to
OCaml~\cite{modular-implicits} ---
 not for implicit instantiation of arguments,
 but because modular implicits support higher-kinded quantification
  with propagation of variance information.
Other approaches to higher-kinded polymorphism could perhaps be used instead
  \cite{higher}. 

This approach to inductive types also makes sense for GADTs, such as
the equality GADT defined below~\cite{
      typing-dynamic-typing, 
      first-class-phantom-types, 
      type-safe-cast
}:
\begin{center}\begin{tabular}{c}\begin{lstlisting}
type ('a, 'b) eq = Refl : ('a, 'a) eq
\end{lstlisting}\end{tabular}\end{center}

This uses OCaml's GADT syntax, but we can do without it by building a Church encoding of equality~\cite{bob-higher-kinds}, using the same technique as before:
\begin{center}\begin{tabular}{c}\begin{lstlisting}
module type EQ = sig
  type ('a, 'b) t
  val refl : ('a, 'a) t
end
type ('a, 'b) eq = {E : EQ} -> ('a, 'b) E.t
\end{lstlisting}\end{tabular}\end{center}

We can use this encoding to implement the standard operations on the
equality GADT by providing a suitable implementation of the
\lstinline{EQ} interface. For instance, we can implement the coercion function
\begin{center}\begin{tabular}{c}\begin{lstlisting}
val cast : ('a, 'b) eq -> 'a -> 'b
\end{lstlisting}\end{tabular}\end{center}
by supplying an
implementation of \lstinline{EQ} using function types:
\begin{center}\begin{tabular}{c}\begin{lstlisting}
let cast (f : ('a, 'b) eq) x =
  let module L = struct
    type ('a, 'b) t = 'a -> 'b
    let refl = fun x -> x
  end in
  f {L} x
\end{lstlisting}\end{tabular}\end{center}

\begin{figure}[t]
\begin{lstlisting}
module type SUB =
sig
  type (-'a, +'b) t
  val refl : ('a, 'a) t
end
  
type (-'a, +'b) sub = {S:SUB} -> ('a,'b) S.t
    
let refl {S:SUB} = S.refl

let lift {P:POS} (f : ('a, 'b) sub) =
  let module L = struct
    type ('a,'b) t = ('a P.t,'b P.t) sub
    let refl = refl
  end in f {L}
  
let (>:) x (f : ('a, 'b) sub) =
  let module L = struct
    type ('a, 'b) t = 'a -> 'b
    let refl = fun x -> x
  end in f {L} x
\end{lstlisting}
\caption{\label{figure:initial-implementation}First-class subtypes as an initial algebra}

\end{figure}

If we modify the signature \lstinline{EQ} by adding co- and
contra-variance markers to the parameters of the type \lstinline{t},
then we get the Church encoding of first-class subtypes, as shown in
Figure~\ref{figure:initial-implementation}, from which we can
implement the \lstinline{refl}, \lstinline{lift} and \lstinline{(:>)}
functions.

\subsection{Converting between encodings}
\label{section:conversion}

Despite the different starting points,
 the three implementations are interdefinable.
In fact, given \textit{any} two implementations \lstinline{A} and
 \lstinline{B} of the subtyping interface, a subtyping witness of type \lstinline{('a,'b) A.t}
 can be converted to a witness of type \lstinline{('a,'b) B.t}.

\noindent
The \lstinline{SUB} module type contains the four elements of 
Figure~\ref{figure:basic-sub-interface}
(\lstinline{t}, \lstinline{refl}, \lstinline{lift}, \lstinline{>:}):

\begin{center}\begin{tabular}{c}\begin{lstlisting}
module type SUB =
sig
  type ('a, 'b) t
  val refl : ('a, 'a) t
  val lift : {P:POS} -> ('a, 'b) t -> ('a P.t, 'b P.t) t
  val (>:) : 'a -> ('a, 'b) sub -> 'b
end
\end{lstlisting}\end{tabular}\end{center}

\noindent
The function \lstinline{conv} takes two implementations of
\lstinline{SUB}, \lstinline{A} and \lstinline{B}, and converts a value
in \lstinline{A.t} to a value of \lstinline{B.t}:

\begin{center}\begin{tabular}{c}\begin{lstlisting}
val conv : {A:SUB} -> {B: SUB} -> ('a, 'b) A.t -> ('a, 'b) B.t
\end{lstlisting}\end{tabular}\end{center}

\noindent
As often, implementing \lstinline{conv} is a matter of finding a
suitable implementation of \lstinline{POS} to pass to \lstinline{lift}:

\begin{center}\begin{tabular}{c}\begin{lstlisting}
let conv (type a b) {A:SUB} {B:SUB} (x : (a,b) A.t) =
  let module M = struct type 'a t = (a, 'a) B.t end in
  A.(>:) B.refl (A.lift {M} x)
\end{lstlisting}\end{tabular}\end{center}

The function \lstinline{conv} works as follows.
The value \lstinline{x} is a proof that \lstinline{a $\le_A$ b} ---
that is, that \lstinline{a} is an \lstinline{A}-subtype of
\lstinline{b}.
The type \lstinline{M.t} represents the positive context \lstinline{a $\le_B$ -}.
The call to \lstinline{lift} then lifts the proof 
\lstinline{a $\le_A$ b} through the context \lstinline{M.t} to produce a proof
\lstinline{(a $\le_B$ a) $\le_A$ (a $\le_B$ b)}.
Finally, this proof can be used to coerce 
  a proof of the reflexivity of \lstinline{B}-subtyping (at type \lstinline{a}) \lstinline{a $\le_B$ a}
  to a proof that \lstinline{a $\le_B$ b}.
That is, from a proof \lstinline{a $\le_A$ b},
the operations of \lstinline{A.sub} and \lstinline{B.sub} produce
a proof \lstinline{a $\le_B$ b}.




\section{First-class subtypes: further examples}
\label{section:further-examples}

\subsection{Arrays and rows}

Here is a function that prints arrays by calling the \lstinline{name}
method of each element:
\begin{center}\begin{tabular}{c}\begin{lstlisting}
  let print_array = Array.iter (fun o -> print o#name)
\end{lstlisting}\end{tabular}\end{center}
To call \lstinline{name}, \lstinline{print_array} does not need to
know the full element type: it needs only to know that there is a
method \lstinline{name} returning \lstinline{string}.
OCaml gives \lstinline{print_array} a row type,
indicating that the element type may have other methods:
\begin{center}\begin{tabular}{c}\begin{lstlisting}
  val print_array : <name: string; ..> array -> unit
\end{lstlisting}\end{tabular}\end{center}
But rows are sometimes too inflexible.
Given two arrays \lstinline{a}, \lstinline{b}, of
different element types
\begin{center}\begin{tabular}{c}\begin{lstlisting}
val a : < m : int; name : string > array
val b : < n : bool; name : string > array
\end{lstlisting}\end{tabular}\end{center}
unification will fail:
\begin{center}\begin{tabular}{c}\begin{lstlisting}[style=text]
List.iter print_array [a; b];;
                          ^
Error: This expression has type < n : bool; name : string > array
       but an expression was expected of type
         < m : int; name : string > array
       The second object type has no method n
\end{lstlisting}\end{tabular}\end{center}
Using first-class subtypes it is possible to combine iterations over
arrays whose element types belong to the same subtyping hierarchy.

\begin{center}\begin{tabular}{c}\begin{lstlisting}
type +'a arr = Arr : 'x array * ('x, 'a) sub -> 'a arr
let aiter f (Arr (a,sub)) = Array.iter (fun s -> f (s >: sub)) a
List.iter
  (aiter (fun o -> print o#name)) [Arr (a,refl); Arr (b,refl)]
\end{lstlisting}\end{tabular}\end{center}







\subsection{Selective abstraction}

A third class of examples arises from selective abstraction, where an
abstract type comes with a proof of a property about that type.
For example, here is a module that exports a type \lstinline{t} along
with a proof that \lstinline{t} is a subtype of \lstinline{int}:
\begin{center}\begin{tabular}{c}\begin{lstlisting}
module M:
  sig type t
  val t_sub_int : (t,int) sub
  (*$\;\ldots\;$*)
end
\end{lstlisting}\end{tabular}\end{center}
Outside the module, values of type \lstinline{t} can be coerced to
\lstinline{int}, but not vice versa.
This approach supports a style similar to refinement types,
 in which abstraction boundaries distinguish values of a type
 for which some additional predicate has been shown to hold.

These are known as \emph{partially abstract types}~\cite{CW-private-types},
and are available as a language feature in OCaml~\cite{private-rows} and Moby~\cite{moby-subtyping}.
However, implementing these via first-class subtypes allows more
flexibility: for example,
  they allow some of the methods of an object type to be
   hidden from the exposed interface,
  and also support the dual of private types
   (called \textit{invisible types}~\cite{gadts-meet-subtyping}),
  and \textit{zero cost-coercions}~\cite{safe-zero-cost-coercions-for-haskell},
   where coercions in both directions are available,
   but actual type equality is not exposed.

\subsection{Bounded quantification}

Dual to abstraction,
combining first-class subtypes with OCaml's
first-class polymorphism encodes bounded quantification.
For example,
 the type
  $\forall \alpha \leq t.\; \alpha \to t$
 might be written as follows:

\begin{center}\begin{tabular}{c}
\begin{lstlisting}
type s = { f: 'a. ('a, t) sub -> 'a -> t }
\end{lstlisting}
\end{tabular}\end{center}

\subsection{Proofs of variance}

Finally, first-class subtypes can express proofs of variance.
For example, the covariance of \lstinline{list}
 can be represented by a value of the following type:

\begin{center}\begin{tabular}{c}
\begin{lstlisting}
('a,'b) sub -> ('a list, 'b list) sub
\end{lstlisting}
\end{tabular}\end{center}

\subsection{Unsoundness in Java}

Amin and Tate~\cite{DBLP:conf/oopsla/AminT16} present an encoding of
first-class subtypes in Java, which they use to demonstrate a
soundness bug. They use Java's bounded quantification to define a type
\lstinline{Constrain<A, B>} which is well-formed only when
\lstinline{A} $\leq$ \lstinline{B}. Then, the type $\exists \texttt{X}
\geq \texttt{T} .\; \texttt{Constrain<} \texttt{U}, \texttt{X}
\texttt{>}$ corresponds to a subtyping witness \lstinline{(U,T) sub}:
if this type is inhabited, then some $\texttt{X} \geq \texttt{T}$ exists making
\lstinline{Constrain<U, X>} well-formed, giving \lstinline{U} $\leq$
\lstinline{X} $\leq$ \lstinline{T}. Unfortunately, in Java (unlike
OCaml), the value \lstinline{null} inhabits every reference type,
giving an invalid subtyping witness that allows any type to be coerced
to any other.

\section{Discussion}

The encodings given here are
 useful for exploratory work,
  for demonstrating soundness,
 and for showcasing OCaml's expressivity.
However, direct language support would make first-class
subtypes more usable.
Scherer and R\'emy~\cite{gadts-meet-subtyping}
discuss design issues and related work
 (e.g.~\cite{
Emir:2006:VGC:2171327.2171352,
             vaugon:tel-01356695}).

The encodings suffer from some awkwardness,
since contexts must be applied explicitly,
  unlike the equalities revealed by pattern matching with \textsc{gadt}s,
     which the type checker applies implicitly.

Our encodings share another issue
  with similar encodings of GADTs~\cite{first-class-modules-ml-2010}:
they lack inversion principles.
Given \lstinline{('a, 'b) sub},
  our encodings can be used to derive \lstinline{('a list, 'b list) sub},
    from the covariance of the \lstinline{list} type constructor.
However, going the other direction,
  from \lstinline{('a list, 'b list) sub} to \lstinline{('a, 'b) sub},
    is equally valid but not expressible with our encodings.

With language support for subtype witnesses,
  coercions would still be explicit,
  but constraints in scope
    could be implicitly lifted through contexts,
      and inversion principles could be applied.



\paragraph*{Acknowledgements}

We thank Fran\c{c}ois Pottier, Leo White, the ML 2017 reviewers, and
the ML \& OCaml 2017 post-proceedings reviewers for helpful comments.

\bibliographystyle{eptcs}
\bibliography{first-class-subtypes}

\begin{thebibliography}{10}
\providecommand{\bibitemdeclare}[2]{}
\providecommand{\surnamestart}{}
\providecommand{\surnameend}{}
\providecommand{\urlprefix}{Available at }
\providecommand{\url}[1]{\texttt{#1}}
\providecommand{\href}[2]{\texttt{#2}}
\providecommand{\urlalt}[2]{\href{#1}{#2}}
\providecommand{\doi}[1]{doi:\urlalt{http://dx.doi.org/#1}{#1}}
\providecommand{\bibinfo}[2]{#2}

\bibitemdeclare{inproceedings}{DBLP:conf/oopsla/AminT16}
\bibitem{DBLP:conf/oopsla/AminT16}
\bibinfo{author}{Nada \surnamestart Amin\surnameend} \& \bibinfo{author}{Ross
  \surnamestart Tate\surnameend} (\bibinfo{year}{2016}):
  \emph{\bibinfo{title}{Java and {S}cala's type systems are unsound: the
  existential crisis of null pointers}}.
\newblock In: {\sl \bibinfo{booktitle}{Proceedings of the 2016 {ACM} {SIGPLAN}
  International Conference on Object-Oriented Programming, Systems, Languages,
  and Applications, {OOPSLA} 2016, part of {SPLASH} 2016, Amsterdam, The
  Netherlands, October 30 - November 4, 2016}}, pp. \bibinfo{pages}{838--848},
  \doi{10.1145/2983990.2984004}.

\bibitemdeclare{inproceedings}{bob-higher-kinds}
\bibitem{bob-higher-kinds}
\bibinfo{author}{Robert \surnamestart Atkey\surnameend} (\bibinfo{year}{2012}):
  \emph{\bibinfo{title}{Relational Parametricity for Higher Kinds}}.
\newblock In \bibinfo{editor}{Patrick \surnamestart C\'{e}gielski\surnameend}
  \& \bibinfo{editor}{Arnaud \surnamestart Durand\surnameend}, editors: {\sl
  \bibinfo{booktitle}{Computer Science Logic (CSL'12)}}, {\sl
  \bibinfo{series}{LIPIcs}}~\bibinfo{volume}{16},
  \doi{10.4230/LIPIcs.CSL.2012.46}.

\bibitemdeclare{inproceedings}{typing-dynamic-typing}
\bibitem{typing-dynamic-typing}
\bibinfo{author}{Arthur~I. \surnamestart Baars\surnameend} \&
  \bibinfo{author}{S.~Doaitse \surnamestart Swierstra\surnameend}
  (\bibinfo{year}{2002}): \emph{\bibinfo{title}{Typing Dynamic Typing}}.
\newblock In: {\sl \bibinfo{booktitle}{Proceedings of the 7th ACM SIGPLAN
  International Conference on Functional Programming}}, \bibinfo{series}{ICFP
  '02}, \bibinfo{publisher}{ACM}, \doi{10.1145/581478.581494}.

\bibitemdeclare{article}{mezzo}
\bibitem{mezzo}
\bibinfo{author}{Thibaut \surnamestart Balabonski\surnameend},
  \bibinfo{author}{Fran{\c{c}}ois \surnamestart Pottier\surnameend} \&
  \bibinfo{author}{Jonathan \surnamestart Protzenko\surnameend}
  (\bibinfo{year}{2016}): \emph{\bibinfo{title}{The Design and Formalization of
  Mezzo, a Permission-Based Programming Language}}.
\newblock {\sl \bibinfo{journal}{{ACM} Trans. Program. Lang. Syst.}}
  \bibinfo{volume}{38}(\bibinfo{number}{4}), pp. \bibinfo{pages}{14:1--14:94},
  \doi{10.1145/2837022}.

\bibitemdeclare{inproceedings}{safe-zero-cost-coercions-for-haskell}
\bibitem{safe-zero-cost-coercions-for-haskell}
\bibinfo{author}{Joachim \surnamestart Breitner\surnameend},
  \bibinfo{author}{Richard~A. \surnamestart Eisenberg\surnameend},
  \bibinfo{author}{Simon \surnamestart Peyton~Jones\surnameend} \&
  \bibinfo{author}{Stephanie \surnamestart Weirich\surnameend}
  (\bibinfo{year}{2014}): \emph{\bibinfo{title}{Safe Zero-cost Coercions for
  Haskell}}.
\newblock In: {\sl \bibinfo{booktitle}{Proceedings of the 19th ACM SIGPLAN
  International Conference on Functional Programming}}, \bibinfo{series}{ICFP
  '14}, \bibinfo{publisher}{ACM}, \doi{10.1145/2628136.2628141}.

\bibitemdeclare{article}{CW-private-types}
\bibitem{CW-private-types}
\bibinfo{author}{Luca \surnamestart Cardelli\surnameend} \&
  \bibinfo{author}{Peter \surnamestart Wegner\surnameend}
  (\bibinfo{year}{1985}): \emph{\bibinfo{title}{On Understanding Types, Data
  Abstraction, and Polymorphism}}.
\newblock {\sl \bibinfo{journal}{ACM Comput. Surv.}}
  \bibinfo{volume}{17}(\bibinfo{number}{4}), pp. \bibinfo{pages}{471--523},
  \doi{10.1145/6041.6042}.

\bibitemdeclare{techreport}{first-class-phantom-types}
\bibitem{first-class-phantom-types}
\bibinfo{author}{James \surnamestart Cheney\surnameend} \&
  \bibinfo{author}{Ralf \surnamestart Hinze\surnameend} (\bibinfo{year}{2003}):
  \emph{\bibinfo{title}{{First-Class Phantom Types}}}.
\newblock \bibinfo{type}{Technical Report}, \bibinfo{institution}{Cornell
  University}.

\bibitemdeclare{article}{church-types}
\bibitem{church-types}
\bibinfo{author}{Alonzo \surnamestart Church\surnameend}
  (\bibinfo{year}{1940}): \emph{\bibinfo{title}{A Formulation of the Simple
  Theory of Types}}.
\newblock {\sl \bibinfo{journal}{The Journal of Symbolic Logic}}
  \bibinfo{volume}{5}(\bibinfo{number}{2}), \doi{10.2307/2266170}.
\newblock \urlprefix\url{http://www.jstor.org/stable/2266170}.

\bibitemdeclare{inproceedings}{cretin:hal-00650910}
\bibitem{cretin:hal-00650910}
\bibinfo{author}{Julien \surnamestart Cretin\surnameend} \&
  \bibinfo{author}{Didier \surnamestart R{\'e}my\surnameend}
  (\bibinfo{year}{2012}): \emph{\bibinfo{title}{{On the Power of Coercion
  Abstraction}}}.
\newblock In: {\sl \bibinfo{booktitle}{{POPL 2012: 39th ACM SIGPLAN-SIGACT
  Symposium on Principle Of Programming Languages}}},
  \bibinfo{series}{Proceedings of the 39th annual ACM SIGPLAN-SIGACT symposium
  on Principles of programming languages}, \bibinfo{organization}{{ACM}},
  \bibinfo{publisher}{{ACM}}, \bibinfo{address}{Philadelphia, United States},
  \doi{10.1145/2103656.2103699}.

\bibitemdeclare{inproceedings}{Emir:2006:VGC:2171327.2171352}
\bibitem{Emir:2006:VGC:2171327.2171352}
\bibinfo{author}{Burak \surnamestart Emir\surnameend}, \bibinfo{author}{Andrew
  \surnamestart Kennedy\surnameend}, \bibinfo{author}{Claudio \surnamestart
  Russo\surnameend} \& \bibinfo{author}{Dachuan \surnamestart Yu\surnameend}
  (\bibinfo{year}{2006}): \emph{\bibinfo{title}{Variance and Generalized
  Constraints for {C\#} Generics}}.
\newblock In: {\sl \bibinfo{booktitle}{Proceedings of the 20th European
  Conference on Object-Oriented Programming}}, \bibinfo{series}{ECOOP'06},
  \bibinfo{publisher}{Springer-Verlag}, \doi{10.1007/11785477_18}.

\bibitemdeclare{inproceedings}{moby-subtyping}
\bibitem{moby-subtyping}
\bibinfo{author}{Kathleen \surnamestart Fisher\surnameend} \&
  \bibinfo{author}{John \surnamestart Reppy\surnameend} (\bibinfo{year}{2000}):
  \emph{\bibinfo{title}{Extending Moby with Inheritance-Based Subtyping}}.
\newblock In \bibinfo{editor}{Elisa \surnamestart Bertino\surnameend}, editor:
  {\sl \bibinfo{booktitle}{ECOOP 2000 --- Object-Oriented Programming}},
  \bibinfo{publisher}{Springer Berlin Heidelberg}, \bibinfo{address}{Berlin,
  Heidelberg}, pp. \bibinfo{pages}{83--107}, \doi{10.1007/3-540-45102-1_5}.

\bibitemdeclare{inbook}{relaxing-the-value-restriction}
\bibitem{relaxing-the-value-restriction}
\bibinfo{author}{Jacques \surnamestart Garrigue\surnameend}
  (\bibinfo{year}{2004}): \emph{\bibinfo{title}{Functional and Logic
  Programming: 7th International Symposium, FLOPS 2004, Nara, Japan, April 7-9,
  2004. Proceedings}}, chapter \bibinfo{chapter}{Relaxing the Value
  Restriction}.
\newblock \bibinfo{publisher}{Springer Berlin Heidelberg}.

\bibitemdeclare{inproceedings}{private-rows}
\bibitem{private-rows}
\bibinfo{author}{Jacques \surnamestart Garrigue\surnameend}
  (\bibinfo{year}{2006}): \emph{\bibinfo{title}{Private Row Types: Abstracting
  the Unnamed}}.
\newblock In \bibinfo{editor}{Naoki \surnamestart Kobayashi\surnameend},
  editor: {\sl \bibinfo{booktitle}{Programming Languages and Systems: 4th Asian
  Symposium, APLAS 2006}}, \bibinfo{publisher}{Springer Berlin Heidelberg},
  \doi{10.1007/11924661_3}.

\bibitemdeclare{misc}{kmett-magpie}
\bibitem{kmett-magpie}
\bibinfo{author}{Edward \surnamestart Kmett\surnameend} (\bibinfo{year}{2010}):
  \emph{\bibinfo{title}{Magpie}}.
\newblock \bibinfo{howpublished}{\url{https://github.com/ekmett/magpie/}. See
  also \url{https://issues.scala-lang.org/browse/SI-4040}}.

\bibitemdeclare{article}{a-behavioural-notion-of-subtyping}
\bibitem{a-behavioural-notion-of-subtyping}
\bibinfo{author}{Barbara~H. \surnamestart Liskov\surnameend} \&
  \bibinfo{author}{Jeannette~M. \surnamestart Wing\surnameend}
  (\bibinfo{year}{1994}): \emph{\bibinfo{title}{A Behavioral Notion of
  Subtyping}}.
\newblock {\sl \bibinfo{journal}{ACM Trans. Program. Lang. Syst.}}
  \bibinfo{volume}{16}(\bibinfo{number}{6}), \doi{10.1145/197320.197383}.

\bibitemdeclare{inproceedings}{gadts-meet-subtyping}
\bibitem{gadts-meet-subtyping}
\bibinfo{author}{Gabriel \surnamestart Scherer\surnameend} \&
  \bibinfo{author}{Didier \surnamestart R{\'{e}}my\surnameend}
  (\bibinfo{year}{2013}): \emph{\bibinfo{title}{{GADT}s Meet Subtyping}}.
\newblock In \bibinfo{editor}{Matthias \surnamestart Felleisen\surnameend} \&
  \bibinfo{editor}{Philippa \surnamestart Gardner\surnameend}, editors: {\sl
  \bibinfo{booktitle}{22nd European Symposium on Programming, {ESOP} 2013}},
  {\sl \bibinfo{series}{Lecture Notes in Computer Science}}
  \bibinfo{volume}{7792}, \bibinfo{publisher}{Springer},
  \doi{10.1007/978-3-642-37036-6_30}.

\bibitemdeclare{phdthesis}{vaugon:tel-01356695}
\bibitem{vaugon:tel-01356695}
\bibinfo{author}{Benoit \surnamestart Vaugon\surnameend}
  (\bibinfo{year}{2016}): \emph{\bibinfo{title}{{Subtyping by Constraint
  Saturation, Theory and Implementation}}}.
\newblock \bibinfo{type}{Theses}, \bibinfo{school}{{Universit{\'e}
  Paris-Saclay}}.
\newblock \urlprefix\url{https://pastel.archives-ouvertes.fr/tel-01356695}.

\bibitemdeclare{article}{propositions-as-types}
\bibitem{propositions-as-types}
\bibinfo{author}{Philip \surnamestart Wadler\surnameend}
  (\bibinfo{year}{2015}): \emph{\bibinfo{title}{Propositions As Types}}.
\newblock {\sl \bibinfo{journal}{Commun. ACM}}
  \bibinfo{volume}{58}(\bibinfo{number}{12}), \doi{10.1145/2699407}.

\bibitemdeclare{article}{type-safe-cast}
\bibitem{type-safe-cast}
\bibinfo{author}{Stephanie \surnamestart Weirich\surnameend}
  (\bibinfo{year}{2004}): \emph{\bibinfo{title}{Functional Pearl: type-safe
  cast}}.
\newblock {\sl \bibinfo{journal}{Journal of Functional Programming}}
  \bibinfo{volume}{14}, \doi{10.1017/S0956796804005179}.

\bibitemdeclare{misc}{modular-implicits}
\bibitem{modular-implicits}
\bibinfo{author}{Leo \surnamestart White\surnameend},
  \bibinfo{author}{Fr\'{e}d\'{e}ric \surnamestart Bour\surnameend} \&
  \bibinfo{author}{Jeremy \surnamestart Yallop\surnameend}
  (\bibinfo{year}{2015}): \emph{\bibinfo{title}{Modular Implicits}}.
\newblock \bibinfo{howpublished}{ACM Workshop on ML 2014 post-proceedings},
  \doi{10.4204/EPTCS.198.2}.

\bibitemdeclare{misc}{first-class-modules-ml-2010}
\bibitem{first-class-modules-ml-2010}
\bibinfo{author}{Jeremy \surnamestart Yallop\surnameend} \&
  \bibinfo{author}{Oleg \surnamestart Kiselyov\surnameend}
  (\bibinfo{year}{2010}): \emph{\bibinfo{title}{First-class modules: hidden
  power and tantalizing promises}}.
\newblock \bibinfo{howpublished}{ACM SIGPLAN Workshop on ML}.
\newblock \bibinfo{note}{Baltimore, Maryland, United States}.

\bibitemdeclare{inproceedings}{higher}
\bibitem{higher}
\bibinfo{author}{Jeremy \surnamestart Yallop\surnameend} \&
  \bibinfo{author}{Leo \surnamestart White\surnameend} (\bibinfo{year}{2014}):
  \emph{\bibinfo{title}{Lightweight Higher-Kinded Polymorphism}}.
\newblock In \bibinfo{editor}{Michael \surnamestart Codish\surnameend} \&
  \bibinfo{editor}{Eijiro \surnamestart Sumii\surnameend}, editors: {\sl
  \bibinfo{booktitle}{Functional and Logic Programming - 12th International
  Symposium, {FLOPS} 2014, Kanazawa, Japan. Proceedings}},
  \doi{10.1007/978-3-319-07151-0_8}.

\end{thebibliography}
\end{document}